# Thickness-dependent Topological Phases and Flat Bands in Rhombohedral Multilayer Graphene

H. B. Xiao[1*], C. Chen[1,2*], X. Sui[3*], S. H. Zhang[4*], M. Z. Sun[1], H. Gao[1], Q. Jiang[1], Q. Li[1,2], L. X. Yang[5], M. Ye[6], F. Y. Zhu[6], M. X. Wang[1], J. P. Liu[1,2], Z. B. Zhang[3], Z. J. Wang[1], Y. L. Chen[1,2,7], K. H. Liu[3,8†], Z. K. Liu[1,2†]

[1]School of Physical Science and Technology, ShanghaiTech University, Shanghai 201210, China

[2]Laboratory for Topological Physics, ShanghaiTech University, Shanghai 201210, China

[3]State Key Laboratory for Mesoscopic Physics, Frontiers Science Centre for Nano-optoelectronics, School of Physics, Peking University, Beijing, China

[4]School of Physics and Electronics, Hunan University, Changsha 410082, People's Republic of China

[5]State Key Laboratory of Low Dimensional Quantum Physics, Department of Physics, Tsinghua University, Beijing, 100084, China

[6]Shanghai Synchrotron Radiation Facility, Shanghai Advanced Research Institute, Chinese Academy of Sciences, Shanghai 201210, P. R. China

[7]Department of Physics, Clarendon Laboratory, University of Oxford, Parks Road, Oxford OX1 3PU, UK

[8]Interdisciplinary Institute of Light-Element Quantum Materials and Research Centre for Light-Element Advanced Materials, Peking University, Beijing, China

*These authors contributed equally to this work

Correspondence to: khliu@pku.edu.cn, liuzhk@shanghaitech.edu.cn

**Rhombohedral multilayer graphene has emerged as an extraordinary platform for investigating exotic quantum states, such as superconductivity and fractional quantum anomalous Hall effects, mainly due to the existence of topological surface flatbands. Despite extensive research efforts, a systematic spectroscopic investigation on the evolution of its electronic structure from thin layers to bulk remains elusive. Using state-of-the-art angle-resolved photoemission spectroscopy with submicron spatial resolution, we directly probe and trace the thickness evolution of the topological electronic structures of rhombohedral multilayer graphene. As the layer number increases, the gapped subbands transform into the 3D Dirac nodes that spirals in the momentum space; while the flatbands are constantly observed around Fermi level, and eventually evolve into the topological drumhead surface states. This unique thickness-dependent topological phase transition can be well captured by the 3D generalization of 1D Su–Schrieffer–Heeger chain in thin layers, to the topological Dirac nodal spiral semimetal in the bulk limit. Our findings establish a solid foundation for exploring the exotic quantum phases with nontrivial topology and correlation effects in rhombohedral multilayer graphene.**

**Introduction**

Rhombohedral multilayer graphene (RMG) has recently been demonstrated as a spectacular research platform hosting rich quantum phenomena. These phenomena include superconductivity [1], various symmetry-broken states (spin, valley, and layer) [2,3], integer and fractional quantum anomalous Hall effect [4,5], and Chen insulator states [6,7]. A critical factor in tuning the properties of RMG is the number of layers, as different quantum phases have been observed only in samples with specific layer counts. For instance, the fractional Chern insulator phase was observed exclusively in a 5-layer sample [5], the quantum anomalous Hall in a 4-layer sample [4], and superconductivity in a 3-layer sample [1].

The intriguing properties of RMG and its layer-tuning capabilities are indebted to its unique electronic structure, characterized by the flat surface bands. The dispersion of these flat bands can be approximately described by $E \sim p^N$ in a two-band model, where $E$ is the kinetic energy, $p$ is the momentum and $N$ is the number of layers [8,9]. As $N$ increases, the flatness of the surface bands increases, leading to strong electron correlations [10,11], which is critical to the realization of many exotic quantum phases in this system, such as superconductivity [1] and fractional quantum anomalous Hall states [4].

The formation of the flat bands could be described by the 3D generalization of the Su–Schrieffer–Heeger (SSH) model [9,12–17], which explains the $2N$-2 gapped bulk subbands and the surface flat bands (SFBs). SFBs correspond to edge states in the SSH model, making RMG 3D generalization of the 1D topological insulator (TI) described by the SSH model. As $N$ increases, both SFBs and subbands can be continuously tuned with the increase of subband numbers and crossing points in the SFB. In the $N = \infty$ limit (rhombohedral graphite), the system is predicted to transition to a topological Dirac nodal spiral semimetal (DNSS), where the gapped subbands merge into a 3D Dirac cone that spirals with opposite helicity around the K and K' points and SFB becomes the drumhead surface state [18].

The intriguing evolution of band structure and topological properties with different $N$, urges a comprehensive investigation on the electronic structure of RMG. To date, several angle-resolved photoemission spectroscopy (ARPES) studies have reported the electronic structure of $N$ = 3 [17], 4 [15], and 14 [14], with limited resolution and comprehensiveness. A systematic investigation of the layer-dependent evolution of the electronic structure, particularly as $N$ approaches ∞ limit, remains elusive. Such studies are essential for understanding the interplay between strong electron correlation, topological phase transitions, and the emergence of exotic quantum states in this system.

In this work, employing synchrotron-based spatially resolved ARPES (NanoARPES) measurements, we systematically explored the electronic structure of RMG for $N$ = 3, 24 and bulk limit ($N$ ~ 50). Our observations reveal the topological electronic structure predicted by the SSH model, including the SFBs and the gapped subbands. As $N$ increases, the topological flat bands are constantly observed, and we also found an increase in the number of the subbands from the bulk states, a reduction of the subband energy spacing, and the closing of the subband gap. For rhombohedral graphite ($N$ ~ 50), the gapped subbands transition into gapless 3D Dirac cones that spiral around K/K' in the momentum space. Meanwhile, the SFBs evolve into drumhead surface states with a large number of crossing points, establishing a topological DNSS system. The observed band evolution is in agreement with the density function theory (DFT) calculations and therefore confirms the thickness-dependent topological phases in RMG. Our observations establish RMG as a unique topological flatband system to investigate strong correlations and topological physics.

**Methods**

*Sample Fabrication:* The rhombohedral graphite was prepared by mechanical exfoliation from our epitaxial graphite grown on single-crystal Ni [19]. The rhombohedral domains of graphene were identified by Raman spectroscopy (WITec alpha300R). The thickness was measured by atomic force microscope (Oxford Cypher S). More information about the sample characterization can be found in Fig. S1.

*Bandstructure calculations:* The DFT calculations are carried out in the framework of the generalized gradient approximation (GGA) functional [20] by employing the Vienna ab initio simulation package [21] with projector augmented wave method [22]. The van der Waals interaction is described by D2 method of Grimme [22] in all calculations.

*ARPES experiment:* The NanoARPES measurements were performed at the BL07U and BL03U endstation of Shanghai Synchrotron Radiation Facility (SSRF) [23]. The base pressure is lower than $1\times10^{-10}$ mbar. The sample was annealed at 160 °C for 3 hours prior to the measurement to desorb absorbates. The beam at BL07U was focused using a Fresnel zone plate and a spatial resolution of ~ 400 nm was achieved. The beam at BL03U was focused using a micro-focusing capillary mirror and a spatial resolution of ~ 4 um was achieved. All the experiments were performed at T = 20 K. The experiment at BL07U was performed using photon energies ranging from 88 to 102 eV with right-handed circular polarized light while the experiment at BL03U was performed using photon energies ranging from 68 eV to 112 eV with linearly horizontal polarized light. Circular polarization is adopted for an enhanced data statistic, and the difference between LH and CR is found minimal (see Fig. S2 for details). The data were collected by a hemispherical Scienta DA30 electron analyzer. The instrument energy and momentum resolution under the measurement conditions were ~10 meV and 0.01 Å, respectively.

*Transmission Electron Microscopy Analysis*: The RMG sample with silica substrate was cut perpendicularly along armchair edge of graphene by focus ion beam technology in Thermal Scientific Helios 5 CX DualBeam. Subsequently, high-resolution transmission electron microscopy and scanning transmission electron microscopy were carried out using a JEOL Grand Arm 300F with double Aberration-correctors. All the TEM/STEM images are taken at 300kV in order to get clear atomic structure of graphene with different stacking. OneView camera binned to 2k × 2k pixel resolution was used to acquire images, which were then post-processed through Gatan Microscopy Suite software.

**Results**

RMG is a graphene allotrope with ABC stacking (Fig. 1a), where each atom connects to a nearest neighbor from an adjacent layer either directly above or below it (see Fig. 1b for the TEM image of the sample we measured). RMG can be conceptually mapped onto a 1D SSH chain, where only the nearest neighbor hopping $\gamma_0$ is taken into account. Just like the SSH chain can host a 1D TI phase with topological bound states localized at its two edges [24], RMG is also topologically nontrivial and hosts SFBs at the top and bottom layers (Fig. 1a), thus could be described as the 3D generalization of the SSH model. More realistically, it is described as the Slonzewski–Weiss–McClure model [9, 25, 26], where $\gamma_0$ to $\gamma_4$ are taken into account, thus creating $N$ crossing points in the SFB, with $N$ number of layers. The total Berry phase is $N\pi$ in the K valley even in the presence of trigonal warping, illustrating the nontrivial topological character of SFBs [8] (Details can be found in Supplementary Information (SI) Text S1). In addition, $N$ crossing points in the SFB and $2N$-2 gapped subbands are formed (Fig. 1d). As $N$ increases, the SFB crossing point number increases and the gap between these subbands gradually closes. In the bulk limit, a topological phase transition into DNSS occurs, where all subbands merge into gapless Dirac cones (Fig. 1c) [18]. The gapless Dirac cones merge with the infinity number of the SFB crossing points, forming a distinctive nodal spiral around the K and K' points in the Brillouin zone (BZ) of RMG (Fig. 1d). The bulk-boundary correspondence of the nodal spiral protects the existence of a drumhead surface states, evolved from the SFBs [27].

Our NanoARPES measurements on $N$ = 3, 24 and 48 RMG confirm the predicted band structure evolution as shown in Fig. 2. From the constant energy contours (Fig. 2a) and the high symmetry cuts across the K/K' point (Fig. 2b(i)(ii)) of $N$ = 3 RMG, we clearly identified the dispersion of the SFBs and two gapped hole-type subbands. The overall band structure shows nice agreement with DFT results (Fig. 2b(iii)). The observed band structures show strong contrast to those in Bernal stacked trilayer (ABA) graphene, where gapless subbands near Dirac points are observed (See Fig. S3). The measured subband gap, estimated from the energy spacing between $E_F$ to the top of hole-type subbands, is about 276 meV. The SFB has a bandwidth of more than 107 meV estimated from $E_F$ to the band bottom, as detailed in the zoomed-in cut in Fig. 2c(i).

In the $N = 24$ sample (nominal thickness, though the actual number of layers may be smaller due to potential stacking faults, see Fig. S4) (Fig. 2d-f), the SFB and the gapped hole-type subbands are still observed. Interestingly, the number of subbands increases with a reduced subband gap (83 meV). The bandwidth of SFB is more than 60 meV, estimated from $E_F$ to the band bottom. It is worth noting that changing the photon energies of the probing beam leads to changes in spectral weight among different subbands (Fig. S5). This demonstrates the development of $k_z$ dispersion, similar to what has been observed in multilayer graphene systems [28].

In the bulk rhombohedral graphite sample ($N = 48$) (Fig. 2g-i), the large number of subbands merge into the bulk continuum, reducing the subband gap to zero and forming the Dirac cone structure. The SFBs evolve into a flat band situated at the Dirac nodes at $E_F$. The measured bandwidth of the SFB is greater than 45 meV as shown in Fig. 2i(iii). The dispersion measured in the bulk sample shows nice agreement with the calculated results (we used $N = 48$ to simulate the bulk situation, see Fig. 2h(iii)). The band parameters with different $N$ are summarized in Table. S1, and additional ARPES data showing the momentum extension of the flat band is presented in Fig. S6.

To further reveal the topological DNSS state in the bulk limit, we perform detailed ARPES investigation on the bulk RMG sample, and the result is presented in Fig. 3. Our photon-energy-dependent measurement first validates the observation of the surface state in the bulk RMG, as a consequence of persistent SFB with increasing number of crossing number with $N$ (see schematic in Fig. 3a). From the measured photoemission spectra at various photon energies (Fig. 3b), we find the band dispersion is composed of two parts. The first one is the bulk Dirac cones exhibiting strong variations, where the intensity of the left side of the cone gradually increases while the right side fades out. Such evolution consists of a counter-clockwise rotation of the Dirac node spiral, moving from the right side to the left side (see schematic on top of each cut in Fig. 3b, also see discussion below). The measured dispersions are well captured by the DFT calculations with different $k_z$ values (bottom row of Fig. 3b), indicating the bulk nature of these bands. The broadening of the observed spectrum is likely due to the $k_z$ broadening effect in photoemission [29], which results in shadow Dirac cones as observed in Fig. 2h(ii) and Fig. 3b. Despite the rapid changes in bulk bands, the flat bands near $E_F$ are constantly observed at different photon energies, which could be better visualized from the stacked plot of the energy distribution curve (EDCs) across the K' point (Fig. 3c). They do not appear in the DFT calculation of the bulk rhombohedral graphite but are captured in the large $N$ slab calculation (shades in the bottom row of Fig. 3b), confirming their surface nature. As the observed surface flat band is surrounded by the Dirac nodal spiral structure, it forms the so-called drumhead surface state. The coexistence of the nodal spiral structure and the drumhead surface state confirms the topological DNSS nature of rhombohedral graphite (See also theoretical model in Text S1).

From the Slonzewski–Weiss–McClure model, at large $N$, the band crossings would be continuously arranged along a helical spiral, forming the Dirac nodal spiral [9, 18]. The helicity of the Dirac node spiral is opposite at K and K' points, due to the time-reversal symmetry. This unique feature can be observed by measuring the Fermi surface (FS) at different $k_z$ values using various photon energies of the incoming beam [30]. From the FS maps probed with different photon energies at K (Fig. 4b) and K' points (Fig. 4c) we can trace the evolution of surface states and the Dirac node by tracing the evolution of the strongest spectral intensity. The observed Dirac nodes are all situated on a circle with the same momentum radius (indicated by the dashed curves in Fig. 4b, c). These nodes rotate clockwise/counterclockwise around K/K' respectively with the increase of $k_z$, forming a spiral structure in the momentum space (see schematic in Fig. 4a, where the arrow indicates the position of Dirac node at each $k_z$). The angle positions of the measured Dirac nodes are plotted as a function of photon energy and $k_z$ values in Fig. 4c. These observations are consistent with theoretical predictions (see Text S1), thus confirming the DNSS nature in bulk rhombohedral graphite.

**Discussion**

While the qualitative evolution trend of subband gap and energy spacing with *N* is consistent with DFT calculations, the flatness of SFBs is tunable with the layer number *N*, suggesting its critical role in forming the unique quantum transport phenomena in RMG. An appropriate correlation level may create the so-called "sweet spot' with optimal band flatness, facilitating the observation of multiple quantum phases [31]. The strong correlation effect manifests in the SFBs near $E_F$, as demonstrated in the various experimental and theoretical works [4,16,32,33]. The increased band flatness at large-N RMG could further enhance electron instabilities, paving the way for interaction-driven correlated quantum states. And the 3D flat band in the bulk RMG holds the 3D quantum geometry of the Bloch wave functions, which leads to a large superfluid stiffness in all three directions [34]. Thus, RMG can serve as a unique platform to study the interplay between topology and correlation.

**Conclusion**

In conclusion, our observation reveals characteristic SFBs and gapped subband electronic structures of RMG, as well as their layer-dependent evolutions. These evolutions demonstrate a unique thickness-driven topological phase transition from a 3D generalization of SSH model to a topological DNSS. Our measurement of the SFBs and nodal spiral structure serves as a cornerstone for further investigation into the electronic structure of RMG, as well as its relation with the nontrivial quantum phases.

**Acknowledgments**

We acknowledge the support from the National Natural Science Foundation of China (92365204, 12274298, 52025023, 12304217), the National Key R&D program of China (Grant No. 2022YFA1604400/03, 2022YFA1403500/04), the Strategic Priority Research Program of Chinese Academy of Sciences (XDB33000000), Fundamental Research Funds for the Central Universities from China, and the New Cornerstone Science Foundation through the XPLORER PRIZE.

Note: After we submitted the manuscript, we note a similar ARPES paper addressing the topological phase of rhombohedral graphite has been published [35].

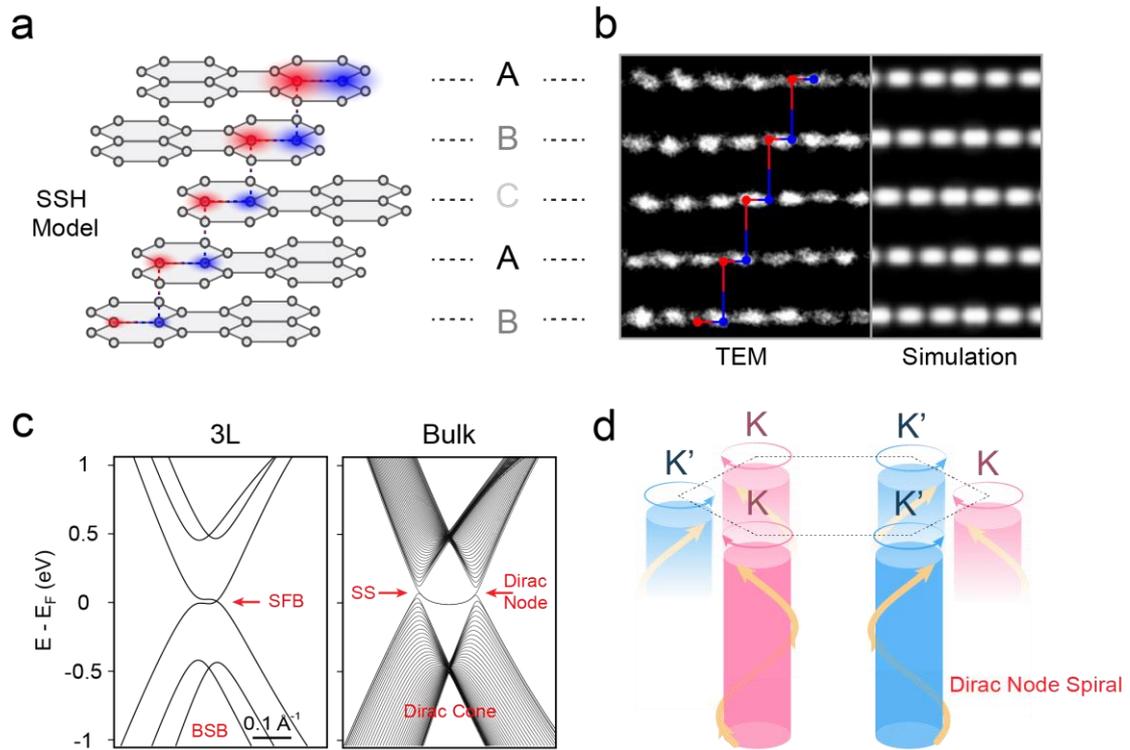

**FIG. 1 Schematic of atomic structure and electronic structure of RMG. a.** Structure of the RMG. Red and blue atoms form the 1D SSH chain, and the colored clouds indicate the wavefunction amplitude located on each atom in the SSH chain. **b.** HRTEM images of RMG, which are consistent with TEM simulation results. Red and blue atoms forming the 1D SSH chain are labelled. **c.** The DFT calculated band structure along the $\Gamma - K - M$ direction for $N$=3 and 100. **d.** Schematic of Dirac node spiral and drumhead surface state in rhombohedral graphite. SFB: surface flat bands. BSB: bulk subbands.

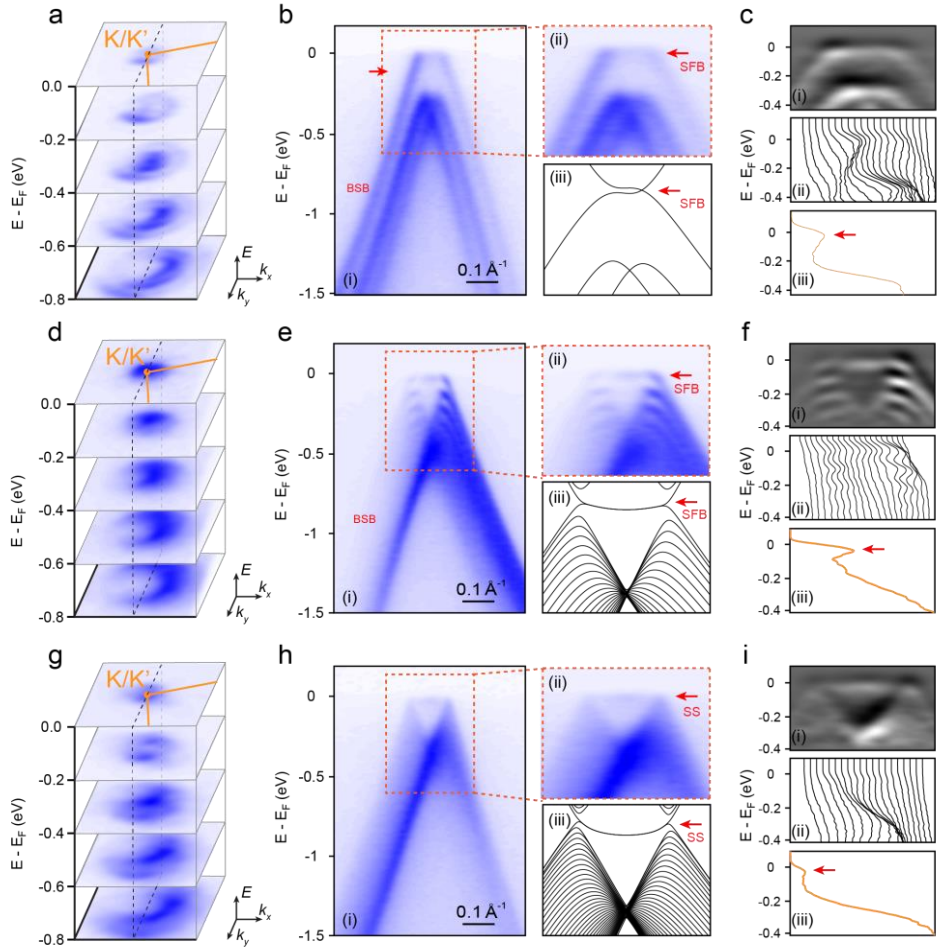

**FIG. 2 Tuning the Electronic structure RMG with layer number *N*. a.** Stacked plot of constant energy contours of *N* = 3 RMG. **b.** (i) Plot of the band dispersion of *N* = 3 RMG along the direction labeled in the dashed line labelled in a. (ii) Zoom in plot around the SFB, marked by red arrow. (iii) DFT calculations of the band structure of *N* = 3 RMG. **c.** (upper) Plot of the second derivative of the photoemission intensity near $E_F$. (middle) Stacked plot of the energy distribution curve of the same region. (lower) Plot of the energy distribution curve at K/K' point. Bandwidth of the SFBs and subband gap sizes are labelled. **d-f**: Same as **a-c** but for *N* = 24 RMG. **g-i**: Same as **a-c** but for *N* = 48 RMG.

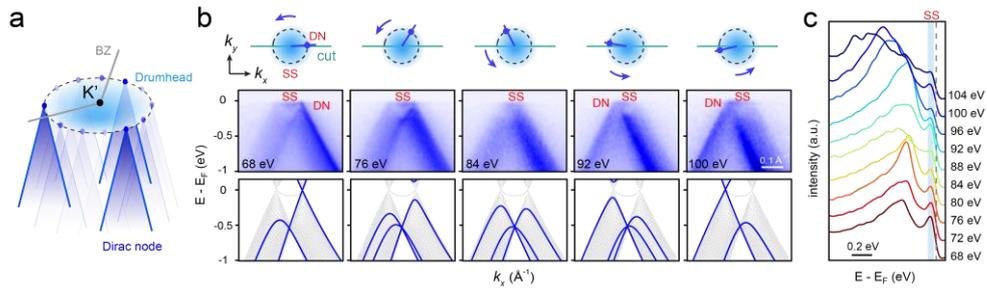

**FIG. 3 Drumhead surface state in rhombohedral graphite. a.** Illustration of Dirac node crossing at the K' point. The number of SFB crossings increases with the increase of RMG layers, forming Dirac nodal lines. The SFB evolves into Drumhead SS. **b.** (Upper) Band dispersion cut across the K' point measured at various photon energies. The cut direction and the relationship between the cut direction and the position of the DN are indicated above each panel. (lower) DFT calculations of the dispersion of rhombohedral graphite along the same direction with the corresponding $k_z$. Shades show the slab calculation results of $N$ = 48 RMG. **c.** Stacked plot of the energy distribution curves along the K' point at different photon energies. The bandwidths of the surface states are labelled by the shade. The data is taken at 20K with linear horizontal polarized photons. SS: surface state. DN: Dirac node.

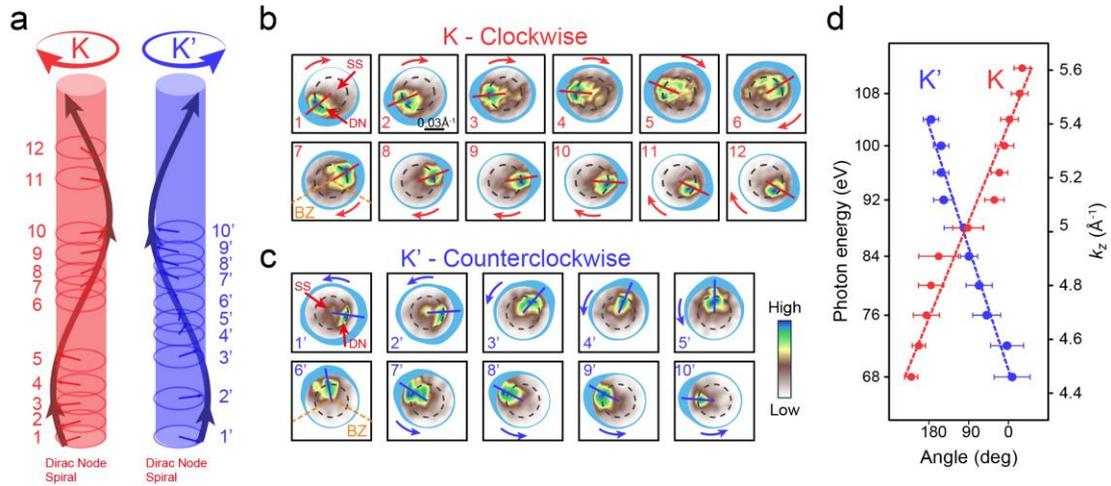

**FIG. 4 Dirac nodal spiral in rhombohedral graphite. a.** Schematic of momentum space Dirac nodal spiral in K and K'. The shaded column indicates the drumhead surface states. The numbered circles indicate the measured Fermi surface (FS) at each photon energy. The arrows indicate the extracted angle of the node measured in **b, c**. **b.** K point FS measured at various photon energies as labelled in **a**. The dashed circle indicates the projection of the spiral. The angular intensity distribution curves were plotted in blue shades outside each map. Red lines indicate the extracted angle of the FS intensity around the dashed circle. The K point is determined by overlaying the Brillouin Zone with respect to the normal direction of the sample (Γ point), and fitting the features of the adjacent K points. **c.** Same as b, but near K' point. **d.** Extracted angle of the Fermi surface intensity around K/K' versus measured photon energies, plotted together with the linear fitting of the helicity. The data is taken at 20K with linear horizontal polarized photons.

**Reference:**

[1] Zhou H, Xie T, Taniguchi T, Watanabe K, Young AF. Superconductivity in rhombohedral trilayer graphene. Nature 2021;598:434–8.

[2] Zhang F, Jung J, Fiete GA, Niu Q, MacDonald AH. Spontaneous Quantum Hall States in Chirally Stacked Few-Layer Graphene Systems. Phys Rev Lett 2011;106:156801.

[3] Liu K, Zheng J, Sha Y, Lyu B, Li F, Park Y, et al. Spontaneous broken-symmetry insulator and metals in tetralayer rhombohedral graphene. Nat Nanotechnol 2024;19:188–95.

[4] Han T, Lu Z, Yao Y, Yang J, Seo J, Yoon C, et al. Large quantum anomalous Hall effect in spin-orbit proximitized rhombohedral graphene. Science 2024;384:647–51.

[5] Lu Z, Han T, Yao Y, Reddy AP, Yang J, Seo J, et al. Fractional quantum anomalous hall effect in multilayer graphene. Nature 2024;626:759–64.

[6] Sha Y, Zheng J, Liu K, Du H, Watanabe K, Taniguchi T, et al. Observation of a chern insulator in crystalline ABCA-tetralayer graphene with spin-orbit coupling. Science 2024;384:414–9.

[7] Han T, Lu Z, Scuri G, Sung J, Wang J, Han T, et al. Correlated insulator and Chern insulators in pentalayer rhombohedral-stacked graphene. Nat Nanotechnol 2024;19:181–7.

[8] Koshino M, McCann E. Trigonal warping and Berry's phase $N\pi$ in ABC-stacked multilayer graphene. Phys Rev B 2009;80:165409.

[9] Zhang F, Sahu B, Min H, MacDonald AH. Band structure of ABC -stacked graphene trilayers. Phys Rev B 2010;82:35409.

[10] Hagymási I, Mohd Isa MS, Tajkov Z, Márity K, Oroszlány L, Koltai J, et al. Observation of competing, correlated ground states in the flat band of rhombohedral graphite. Science Advances 2022;8:eabo6879.

[11] Kerelsky A, Rubio-Verdú C, Xian L, Kennes DM, Halbertal D, Finney N, et al. Moiréless correlations in ABCA graphene. Proc Natl Acad Sci 2021;118:e2017366118.

[12] Su WP, Schrieffer JR, Heeger AJ. Solitons in Polyacetylene. Phys Rev Lett 1979;42:1698–701.

[13] Min H, MacDonald AH. Electronic structure of multilayer graphene. Prog Theor Phys 2008;176:227–52.

[14] Henck H, Avila J, Ben Aziza Z, Pierucci D, Baima J, Pamuk B, et al. Flat electronic bands in long sequences of rhombohedral-stacked graphene. Phys Rev B 2018;97:245421.

[15] Wang W, Shi Y, Zakharov AA, Syväjärvi M, Yakimova R, Uhrberg RIG, et al. Flat-Band Electronic Structure and Interlayer Spacing Influence in Rhombohedral Four-Layer Graphene. Nano Lett 2018;18:5862–6.

[16] Zhou H, Xie T, Ghazaryan A, Holder T, Ehrets JR, Spanton EM, et al. Half- and quarter-metals in rhombohedral trilayer graphene. Nature 2021;598:429–33.

[17] Bao C, Yao W, Wang E, Chen C, Avila J, Asensio MC, et al. Stacking-Dependent Electronic Structure of Trilayer Graphene Resolved by Nanospot Angle-Resolved Photoemission Spectroscopy. Nano Lett 2017;17:1564–8.

[18] Ho C-H, Chang C-P, Lin M-F. Evolution and dimensional crossover from the bulk subbands in ABC-stacked graphene to a three-dimensional Dirac cone structure in rhombohedral graphite. Phys Rev B 2016;93:075437.

[19] Zhang Z, Ding M, Cheng T, Qiao R, Zhao M, Luo M, et al. Continuous epitaxy of single-crystal graphite films by isothermal carbon diffusion through nickel. Nat Nanotechnol 2022;17:1258–64.

[20] Perdew JP, Burke K, Ernzerhof M. Generalized Gradient Approximation Made Simple. Phys Rev Lett 1996;77:3865–8.

[21] Kresse G, Hafner J. *Ab initio* molecular dynamics for liquid metals. Phys Rev B 1993;47:558–61.

[22] Blöchl PE. Projector augmented-wave method. Phys Rev B 1994;50:17953–79.

[23] Gao H, Xiao H, Wang F, Zhu F, Wang M, Liu Z, et al. Nano-ARPES Endstation at BL07U of Shanghai Synchrotron Radiation Facility. Synchrotron Radiat News 2024;37:12–7.

[24] Asbóth JK, Oroszlány L, Pályi A. A Short Course on Topological Insulators. vol. 919. Cham: Springer International Publishing; 2016.

[25] Slonczewski JC, Weiss PR, Band Structure of Graphite. Phys. Rev. 1958;109:272.

[26] McClure W, Band structure of graphite and the de-Haas-van Alphen effect. Phys. Rev. 1957;08:612.

[27] Xiao R, Tasnádi F, Koepernik K, Venderbos JWF, Richter M, Taut M. Density functional investigation of rhombohedral stacks of graphene: Topological surface states, nonlinear dielectric response, and bulk limit. Phys Rev B 2011;84:165404.

[28] Ohta T, Bostwick A, McChesney JL, Seyller T, Horn K, Rotenberg E, Interlayer Interaction and Electronic Screening in Multilayer Graphene Investigated with Angle-Resolved Photoemission Spectroscopy. Phys Rev Lett 2007;98:206802.

[29] Strocov VN, Intrinsic accuracy in 3-dimensional photoemission band mapping. Journal of Electron Spectroscopy and Related Phenomena 2003;130:65-78

[30] Sobota JA, He Y, Shen Z-X. Angle-resolved photoemission studies of quantum materials. Rev Mod Phys 2021;93:025006.

[31] Cao Y. Rhombohedral graphene goes correlated at four or five layers. Nat Nanotechnol 2024;19:139–40.

[32] Shi Y, Xu S, Yang Y, Slizovskiy S, Morozov SV, Son S-K, et al. Electronic phase separation in multilayer rhombohedral graphite. Nature 2020;584:210–4.

[33] Zhou Y-Y, Zhang Y, Zhang S, Cai H, Tong L-H, Tian Y, et al. Layer-dependent evolution of electronic structures and correlations in rhombohedral multilayer graphene arXiv:2312.13637.

[34] Lau A, Hyart T, Autieri C, Chen A, Pikulin DI. Designing Three-Dimensional Flat Bands in Nodal-Line Semimetals. Phys Rev X 2021;11:031017.

[35] Zhang HY, Li Q, Scheer MG, Wang RQ, Tuo CY, Zou NL, et al. Correlated topological flat bands in rhombohedral graphite, PNAS 2024; 121: e2410714121